\documentclass[aps,prb,twocolumn,superscriptaddress,showpacs,amsmath,amssymb,
amsfonts,balancelastpage]{revtex4-2}
\usepackage{amsmath}
\usepackage{amssymb}
\usepackage{amsfonts}
\usepackage[justification=raggedright]{caption}
\usepackage{subcaption}
\usepackage{graphicx}
\usepackage{epsfig}
\usepackage{hyperref}
\usepackage{epstopdf}
\usepackage{gensymb}
\usepackage{color }
\usepackage{color, soul }
\usepackage{xcolor}
\hypersetup{
	colorlinks=true,
	urlcolor=blue,
	citecolor=blue,
	linkcolor=blue,}
\begin{document}
		\title{ Unusual Magnetotransport from two dimensional Dirac Fermions in Pd$_{3}$Bi$_{2}$Se$_{2}$}
	
	\author{Shama}
	\affiliation{Department of Physical Sciences, Indian Institute of Science Education and Research, Knowledge city, Sector 81, SAS Nagar, Manauli PO 140306, Mohali, Punjab, India}
	
    \author{Dinesh Dixit}
	\affiliation{Central Research Facility, Indian Institute of Technology Delhi, Hauz Khas, New Delhi, 110016}
	
	\author{Goutam Sheet}
	\affiliation{Department of Physical Sciences, Indian Institute of Science Education and Research, Knowledge city, Sector 81, SAS Nagar, Manauli PO 140306, Mohali, Punjab, India}
	
	\author{Yogesh Singh} \email{yogesh@iisermohali.ac.in}
	\affiliation{Department of Physical Sciences, Indian Institute of Science Education and Research, Knowledge city, Sector 81, SAS Nagar, Manauli PO 140306, Mohali, Punjab, India}
	
	\begin{abstract}
Pd$_{3}$Bi$_{2}$Se$_{2}$ has been proposed to be topologically non-trivial in nature. However, evidence of its non-trivial behavior is still unexplored. We report the growth and magneto-transport study of Pd$_{3}$Bi$_{2}$Se$_{2}$ thin films, revealing for the first time the contribution of two-dimensional (2D) topological surface states.  We observe exceptional non-saturated linear magnetoresistance which results from Dirac fermions inhabiting the lowest Landau level in the quantum limit. The transverse magnetoresistance changes from a semi-classical weak-field $B^{2}$ dependence to a high-field $B$ dependence at a critical field $B^{\star}$. It is found that $B^{\star} \propto T^2$, which is expected from the Landau level splitting of a linear energy dispersion. In addition, the magnetoconductivity shows signatures of 2D weak anti-localization (WAL). These novel magnetotransport signatures evince the presence of 2D Dirac fermions in Pd$_{3}$Bi$_{2}$Se$_{2}$ thin films.

	\end{abstract}
		\maketitle
	
	\section{Introduction}
	Rapid recent progress has been made in the search for topological materials hosting exotic fermions.\cite{Hasan,Moore,Burkov,Wan,Armitage} Topological semimetals (TS) comprise a large family of materials exhibiting degenerate band crossings near the Fermi level\cite{Burkov,Wan,Armitage}, such as Weyl\cite{Alidoust,Belopolski,Kushwaha}, Dirac\cite{Young,Wang,Weng,Liu}, and multi-fold fermions\cite{Wieder,Vergniory}. In a Dirac semimetal (DSM), the conduction and valence bands touch only at discrete (Dirac) points in the Brillouin zone (BZ) and disperse linearly in all three momentum directions.\cite{Armitage,Young,Wang,Weng,Liu} The Dirac nodes are protected by time-reversal symmetry (TRS) and inversion symmetry (IS). The DSM can be driven into other exotic topological phases including the Weyl semimetal (WSM), by breaking either time-reversal or inversion symmetry.\cite{Armitage,Alidoust,Belopolski,Kushwaha} The topological semimetals exhibit several novel properties such as extremely large magneto-resistance, chiral anomaly induced negative magnetoresistance, extremely high mobility, anomalous Hall effect, and planar Hall effect.\cite{Zhang,Xiong,Shekhar,Liang}
	
	Recently, a family of ternary compounds (parkerite and shandite) with the chemical formula T$_{3}$M$_{2}$X$_{2}$ (T = Ni, Co, Rh, Pd or Pt; M = In, Sn, Pb, Tl or Bi; and X = S, Se or Te) has attracted a lot of interest.\cite{Sakamoto,Wakeshima,Zakharova} The T$_{3}$M$_{2}$X$_{2}$ family of compounds such as Pt$_{3}$Bi$_{2}$Se$_{2}$, Ni$_{3}$Bi$_{2}$Se$_{2}$, Pd$_{3}$Bi$_{2}$Se$_{2}$, and Rh$_{3}$Bi$_{2}$Se$_{2}$ manifest superconductivity and charge density wave behavior in bulk single crystals.\cite{Sakamoto,Wakeshima,Zakharova,Alessandro} Among the members of this material family, being host to non-trivial topological behavior, the materials Co$_{3}$Sn$_{2}$S$_{2}$ and Pd$_{3}$Bi$_{2}$S$_{2}$ have been most
	widely studied.\cite{Sun,Lou,Singh,Roy,Shama} Another member Pd$_{3}$Bi$_{2}$Se$_{2}$ has also been proposed to be topologically non-trivial in nature.\cite{Cano,Po} Previous studies have revealed superconductivity below $1$~K in bulk, thin films, and nanoparticles of Pd$_{3}$Bi$_{2}$Se$_{2}$ .\cite{Wakeshima,Alessandro,Roslova} However, evidence for the non-trivial topological behaviour of Pd$_{3}$Bi$_{2}$Se$_{2}$ is still unexplored.

   It is important to identify signatures of the non-trivial topology in easily accessible measurements like magneto-transport. Unlike the conventional electron gas with parabolic dispersion\cite{Abrikosov}, topological materials have a linear energy dispersion that leads to a large separation between the lowest and the first Landau level (LL) of Dirac fermions.\cite{Jiang,Miller} Consequently, quantum Hall effect, unsaturated linear magnetoresistance, and non-trivial Berry phase could be observed in Dirac semimetal candidates.\cite{Qu,Huynh} Additionally, in magneto-transport measurements, the electron phase can be probed via the quantum (constructive or destructive) interference of electron waves that occurs between electrons traveling through time-reversed paths.\cite{Bergmann,Hikami,Fang,Lee} For materials with weak spin-orbit coupling, constructive interference occurs, which leads to weak localization (WL). The presence of a Berry curvature in momentum space may lead to an extra phase shift of $\pi$ for such closed trajectories, resulting in weak antilocalization (WAL).\cite{Roushan,Ostrovsky,Lu,He} This phase shift is a direct consequence of spin-momentum locking. The magnetic field breaks time-reversal symmetry, which is required for interference. Thus quantum (constructive/destructive) interference can be manifested as a positive (negative) magnetoconductance in low magnetic fields.
  	
In this article, we explore the magneto-transport properties of  Pd$_{3}$Bi$_{2}$Se$_{2}$ thin films grown by pulsed laser deposition technique (PLD). We observe signatures of a charge density wave like transition in temperature dependent resistance measurements. The predicted linear energy dispersion leads to several novel features in transport measurements. We observe exceptional non-saturated linear magnetoresistance when all Dirac fermions inhabit the lowest Landau level in the quantum limit. At a critical field B$^{\star}$, this transverse magnetoresistance shifts from semi-classical weak-field B$^{2}$ dependence to a high-field B dependency. The critical field B$^{\star}$ follows a quadratic temperature dependence, which is expected from the Landau level splitting of linear energy dispersion. Additionally, the magneto-conductance has a sharp cusp-like behavior at low temperatures, which is a hallmark of weak anti-localization (WAL). Applying the HLN theory, we extract the dephasing length (L$_{\phi}$) and $\alpha$, which suggests the presence of multiple topological conduction channels. It is also found that in addition to electron-electron scattering, electron‑phonon scattering also contributes to the phase relaxation mechanism in Pd$_{3}$Bi$_{2}$Se$_{2}$ thin films.\cite{Bird,Sergeev,Reizer}  These novel magnetotransport signatures evince the existence of 2D Dirac fermions in Pd$_{3}$Bi$_{2}$Se$_{2}$ thin films.

\begin{figure}
	\centering
	\includegraphics[width=0.9\linewidth, height=0.25\textheight]{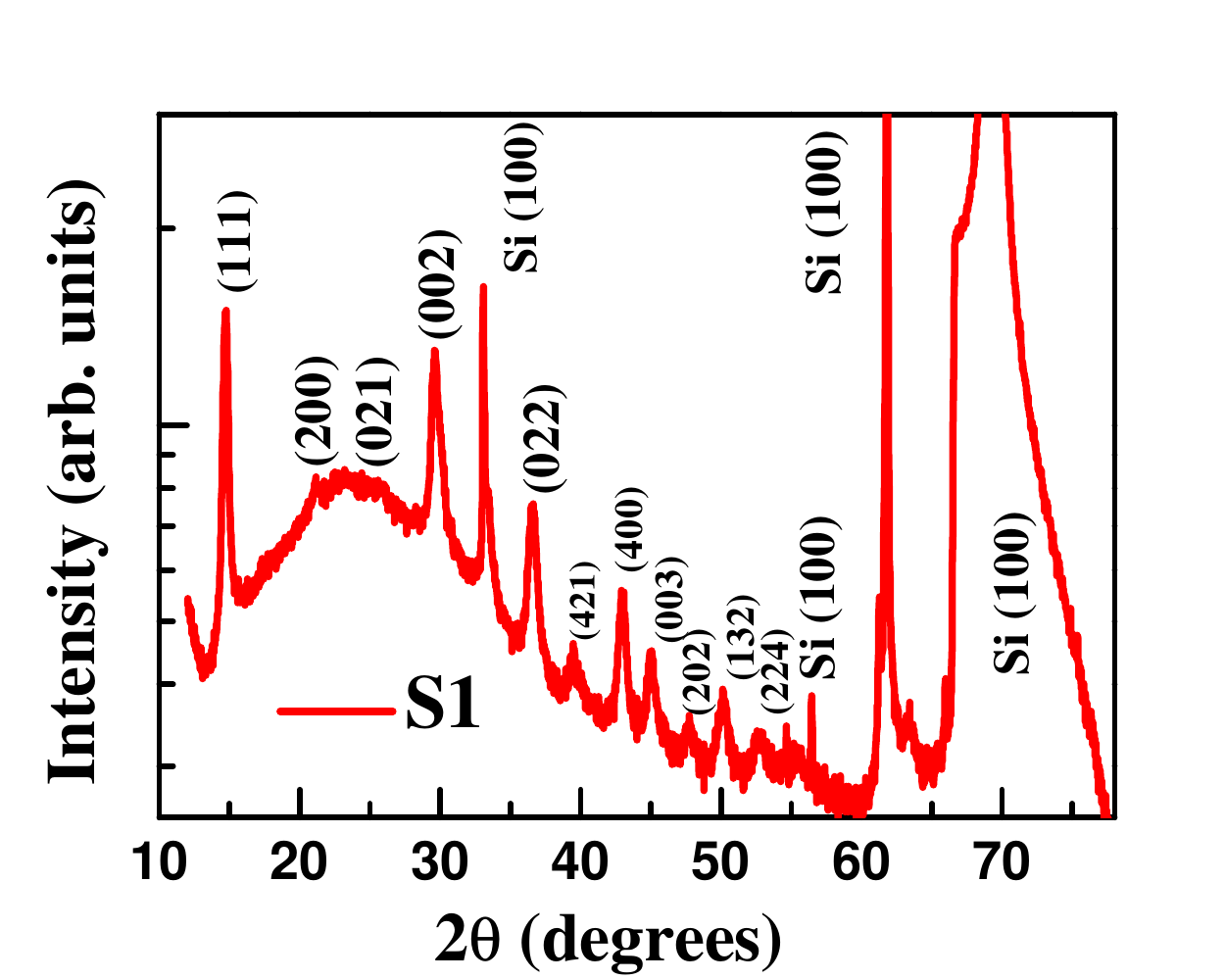}
	\caption{X-ray diffraction data for grown S1 thin films of Pd$_{3}$Bi$_{2}$Se$_{2}$}
	\label{fig:xrd}
\end{figure}

\section{Methods}
Pd$_{3}$Bi$_{2}$Se$_{2}$ thin films were grown using a pulsed laser deposition technique following the procedure described previously for Pd$_3$Bi$_2$S$_2$ \cite{Shama}. The single crystalline substrate (Si 100) was kept at 240$^{\circ}$C during the growth of the films. An x-ray diffractometer (Bruker D8 Advance system with Cu-K$\alpha$ radiation) was used to determine the phase purity of Pd$_{3}$Bi$_{2}$Se$_{2}$ thin-films. The stoichiometry of PBS thin film was confirmed using energy dispersive spectroscopy with a scanning electron microscope (SEM). Pd$_{3}$Bi$_{2}$Se$_{2}$ thin films with different thicknesses (50nm, 22nm) are designated as S1, S2, respectively. The transport measurements were performed in a physical property measurement system ( CFMS 14T, Cryogenics Ltd.).  
 
\section{Results and Discussion}
\begin{figure}
	\centering
	\includegraphics[width=1\linewidth, height=0.3\textheight]{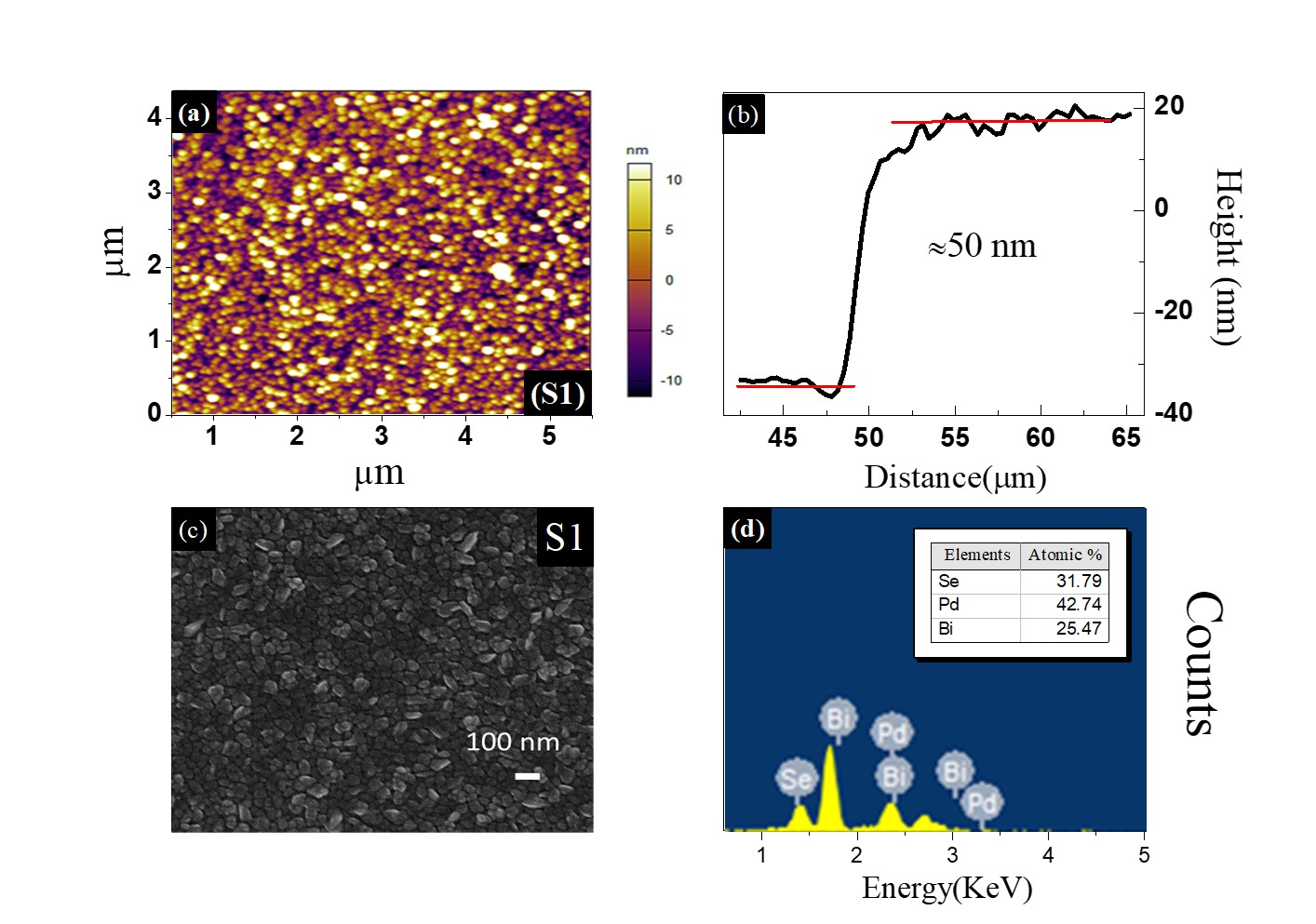}
	\caption{The atomic force microscope topography image of the surface of S1 Pd$_{3}$Bi$_{2}$Se$_{2}$ thin film. (b) The height profile across S1 giving a thickness $\approx$ 50 nm. (c) An SEM image of the surface of S1. (d) Results of EDS spectroscopy on S1 showing the presence of Pd, Bi, Se in the stoichiometric amounts.}
	\label{fig:afm}
\end{figure}

Figure~\ref{fig:xrd} shows the x-ray diffraction (XRD) pattern of Pd$_{3}$Bi$_{2}$Se$_{2}$ thin films measured at T= 300 K, which matches well with the theoretical pattern. This demonstrates the phase purity of the grown thin films. \cite{Wakeshima,Alessandro,Roslova} Figure~\ref{fig:afm}(a) shows the atomic force microscope (AFM) 
topography image of Pd$_{3}$Bi$_{2}$Se$_{2}$ thin films. Using the AFM we measured the roughness and thickness of the sample, which is 50 nm for S1 films, as shown in Fig.~\ref{fig:afm}(b). Figure~\ref{fig:afm}(c) shows the SEM image of the S1 sample. Figure~\ref{fig:afm}(d) shows energy dispersive X-ray spectroscopy data, where peaks of bismuth, palladium, and selenium are observed. The atomic ratio of these elements confirms the stoichiometry of the thin films. These characterization data demonstrate the growth of polycrystalline Pd$_{3}$Bi$_{2}$Se$_{2}$ thin films.

\begin{figure*}
	\centering
	\includegraphics[width=.95\linewidth, height=0.5\textheight]{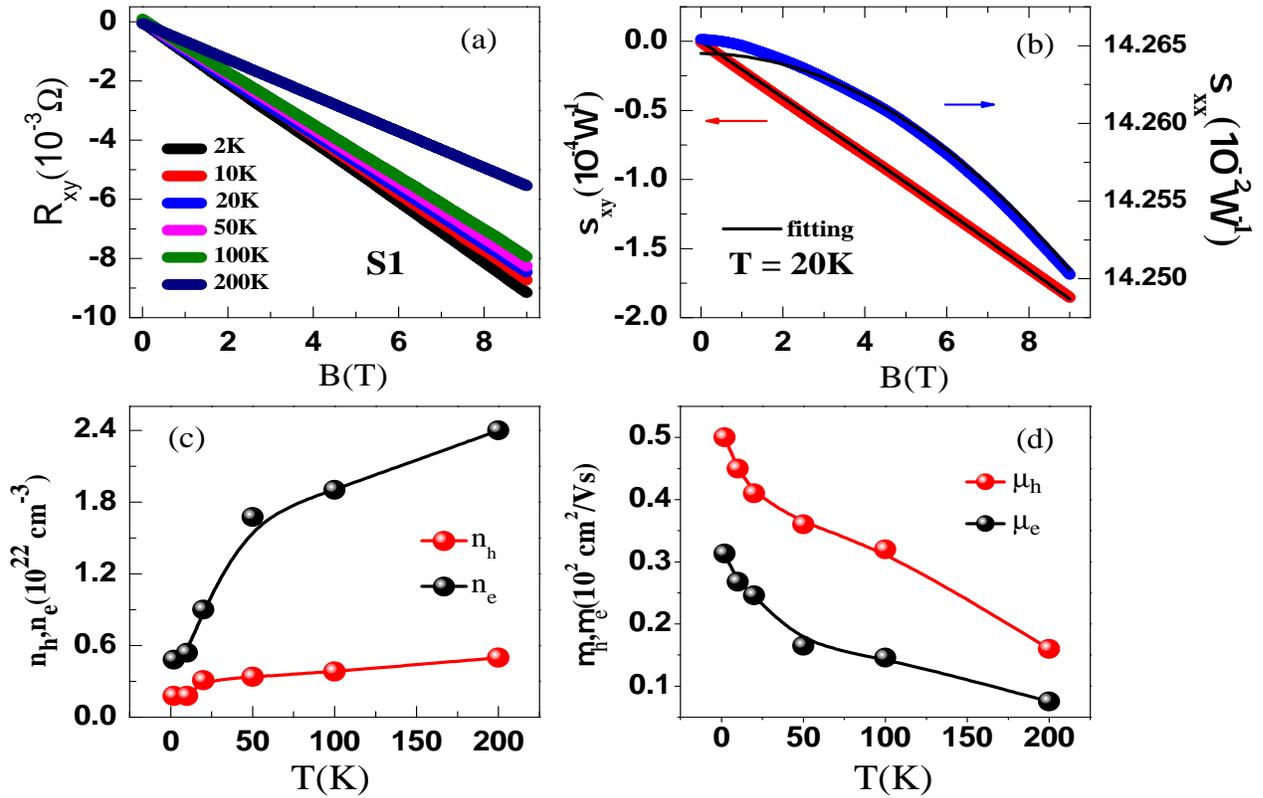}
	\caption{ The Hall resistance (R$_{xy}$) vs magnetic Field at various temperatures for S1 thin film of Pd$_{3}$Bi$_{2}$Se$_{2}$. (b) Variation Hall conductance ($\sigma$$_{xy}$) and longitudinal conductance ($\sigma$$_{xx}$) vs magnetic field at 20K. The solid curves through the data are the global fitting of $\sigma$$_{xy}$ and $\sigma$$_{xx}$ by two band model. (c) The temperature dependence of carrier density and (d) mobility for electron and hole carriers in S1.  }
	\label{fig:hall}
\end{figure*}

To determine the carrier density and its mobility, the Hall resistivity (R$_{xy}$) was measured. Figure~\ref{fig:hall}(a) shows the variation of Hall resistivity (R$_{xy}$) with magnetic field (B) at various temperatures for S1. At high temperature, R$_{xy}$ is almost linear in the field and is negative. However, with decreasing temperature,
R$_{xy}$(B) becomes non-linear, which implies the presence of more than one type of charge carrier. To have more insight, we fit simultaneously and globally the Hall conductance ($\sigma$$_{xy}$) and longitudinal conductance ($\sigma$$_{xx}$) by a semi-classical two-band model, \cite{Namrata,Hurd}
\begin{equation}\label{key}
\sigma_{xy} = e B \Bigg[ \dfrac{n_{h}\mu_{h}^{2}}{1 + (\mu_{h}B)^{2}} - \dfrac{n_{e}\mu_{e}^{2}}{1 + (\mu_{e}B)^{2}}\Bigg]
\end{equation}
\begin{equation}\label{key}
\sigma_{xx} = e \Bigg[ \dfrac{n_{h}\mu_{h}}{1 + (\mu_{h}B)^{2}} + \dfrac{n_{e}\mu_{e}}{1 + (\mu_{e}B)^{2}}\Bigg]
\end{equation}
where e is the charge of the electron and B is the magnetic field. The n$_{h(e)}$ and $\mu_{h(e)}$ are the sheet carrier density and mobility, respectively. The subscript e, h denotes electrons and holes, respectively. Figure~\ref{fig:hall}(b) shows the fit of Hall conductance ($\sigma_{xy}$) and longitudinal conductance ($\sigma_{xx}$) by solid curves. The extracted values of carrier density are of the order 10$^{22}$/cm$^3$ and mobility is 10$^{2}$~cm$^2$/V~s, which are close to the values reported previously.\cite{Alessandro} The extracted parameter rules out the possibility of charge carrier compensation and electrons seem to be the dominant charge carriers.  The temperature dependence of $n$ and $\mu$ are shown in Figs.~\ref{fig:hall}(c) and (d).

\begin{figure}
	\centering
	\includegraphics[width=1.1\linewidth]{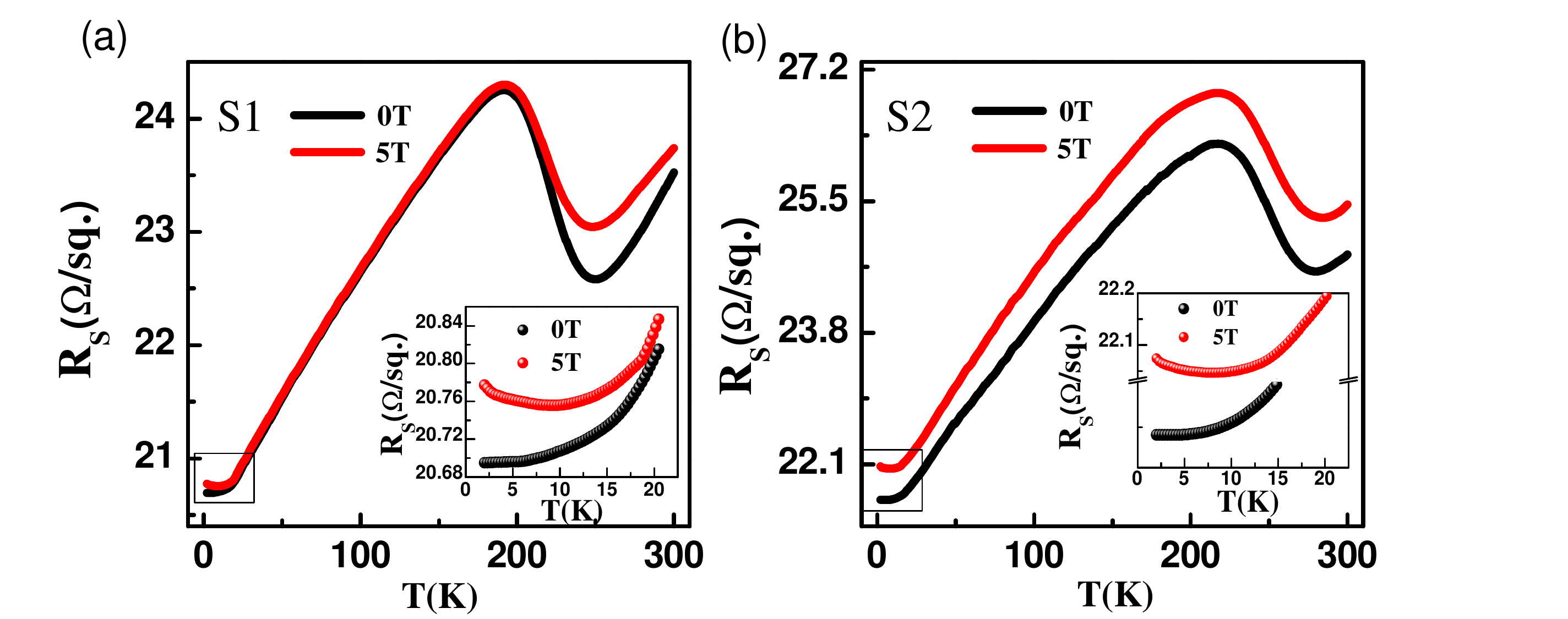}
	\caption{Sheet resistance R$_S$ vs temperature at various magnetic fields showing the semi-conducting  behavior of (a) S1 and (b) S2 films. Insets show an upturn in R$_S$ at low temperatures.}
	\label{fig:rt}
\end{figure}

Figure~\ref{fig:rt}(a-b) shows the variation of sheet resistance (R$_{S}$) with temperature for S1 and S2 films in magnetic fields B = 0T, 5T. We observe metallic behavior with a significant residual resistance R$_{o}$ at the lowest temperature, indicating the presence of some disorder. Under sufficiently strong magnetic field, R$_{S}$(T) below $\sim 10$~K shows an upturn signalling a tendency towards insulating behaviour as can be seen in Fig.~\ref{fig:rt}~(a) and (b) insets.  The temperature at which the minimum in R$_S$(T) occurs increases with $B$. Such an upturn in resistance on the application of a field has been observed in several topological materials although an understanding of its origin has not yet been established.\cite{Hyunwoo}  

Another notable feature in the R$_S$(T) is an abrupt upturn in resistance below $250$~K for S1 and $275$~K for S2 as seen in Fig.~\ref{fig:rt}(a) and (b).  Below these temperatures the resistance first increases, goes over a maximum, and then continues in a metallic fashion for lower temperatures.  This behaviour is similar to that observed for several charge density wave (CDW) materials.\cite{YogeshCDW1, YogeshCDW2} These CDW like features were not observed in the bulk sample reported previously.\cite{Wakeshima}  However we note that Rh$_3$Bi$_2$Se$_2$ a member of the same family, has been reported to show CDW like features in transport measurements \cite{Sakamoto}.  We currently do not have an understanding of the origin of this anomaly in thin films of Pd$_{3}$Bi$_{2}$Se$_{2}$ but we speculate that the enhanced quasi-two-dimensionality may lead to nesting features in the electronic structure which may lead to CDW like instabilities.

\begin{figure}
\centering
\includegraphics[width=1\linewidth]{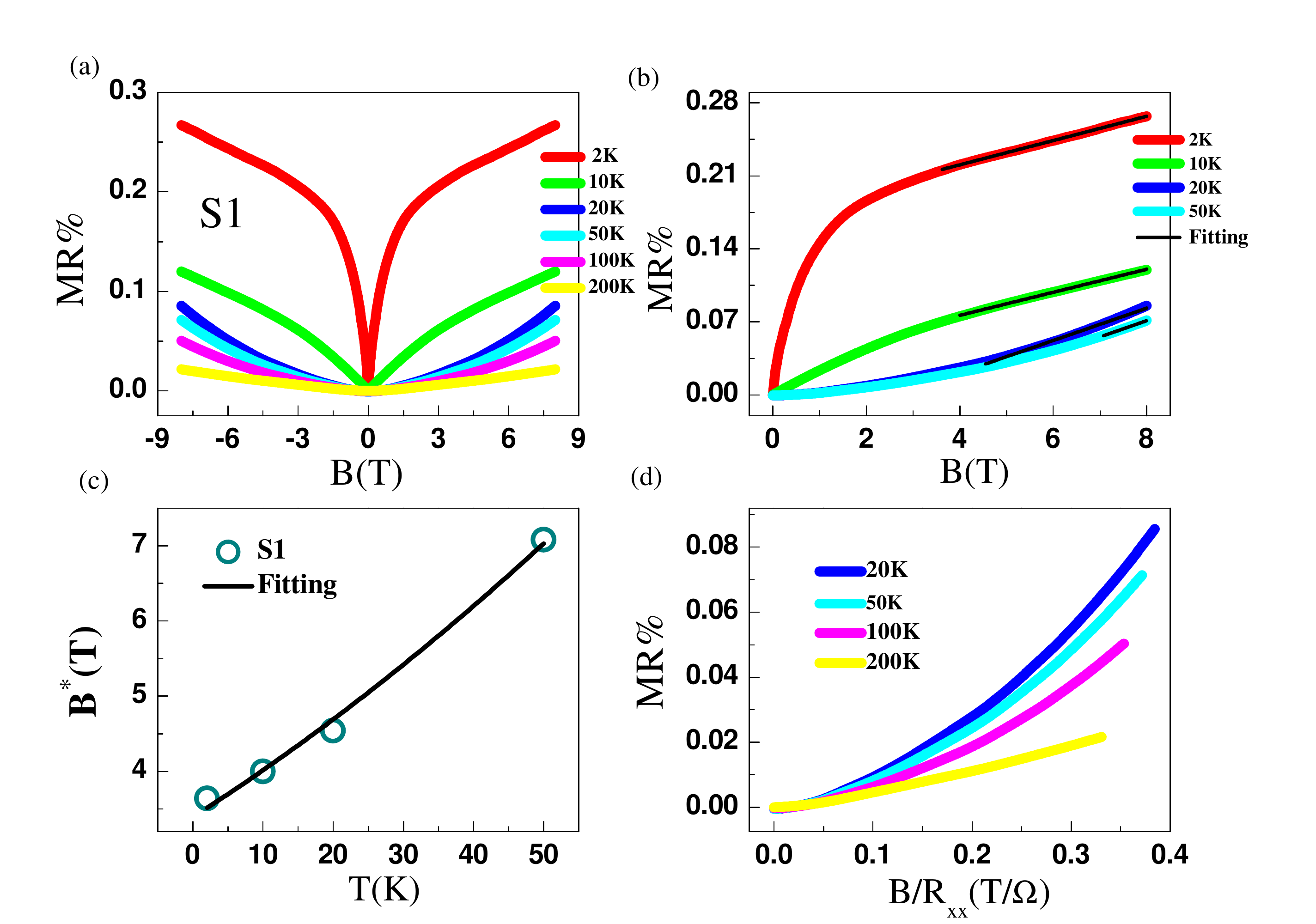}
\caption{(a) Magneto-resistance (MR\%) vs magnetic field (B) at various temperatures in transverse configuration for S1 sample. (b) MR\% vs B at various temperatures. The black solid lines are linear fits in the high magnetic field region. (c) Temperature variation of critical magnetic field B$^{\star}$. (d) The violation of Kohler's plot indicating the presence multiple scattering mechanism.}
\label{fig:mr}
\end{figure}

In Fig.~\ref{fig:mr}(a), the percentage magnetoresistance (MR\% = $\dfrac{R(B) - R(0)}{R(0)}$ $\times$ 100) is plotted as a function of B at various temperatures for B applied perpendicular to the plane of the film. At low temperatures, we observe a logarithmic cusp around zero magnetic
fields, which is a signature of WAL.\cite{Bergmann,Hikami,Fang,Lee} Increasing the magnetic field results in a non-saturating linear response at higher fields. As T increases, the linear dependence of MR turns into the classical $B^2$ dependence, as shown in Fig.~\ref{fig:mr}(b). The MR of bands with a parabolic dispersion either saturates at high fields or increases as B$^{2}$. Such an unusual non-saturating
linear magnetoresistance has previously been reported in the gapless semiconductor Ag$_{2-\delta}$(Te/Se) and was understood to arise due to a linear energy spectrum in the quantum limit.\cite{Husmann,Rosenbaum} Recently, first-principle calculations confirmed a gapless Dirac-type surface state in these materials.\cite{Feng} Linear magnetoresistance is also observed in Bi$_{2}$Te$_{3}$, BaFe$_{2}$As$_{2}$, and (Ca, Sr)MnBi$_{2}$ with Dirac fermion states.\cite{Qu,Huynh,Graf,Hechang} Applying a magnetic
field leads to a quantization of the orbital
motion of carriers with linear energy dispersion, which results in quantized Landau levels (LLs). The energy splitting between the lowest and first LLs
is described by $\triangle_{LL}$ = $\pm$ v$_{F}$ $\sqrt{2e \hbar B}$, where v$_{F}$ is the Fermi velocity, E$_{F}$ is the Fermi energy.\cite{Jiang,Miller} 
In the quantum limit at a specific value of magnetic field and temperature, $\triangle_{LL}$ becomes larger than
the Fermi energy E$_{F}$ and the thermal fluctuations k$_{B}$T
at a finite temperature. Consequently, all carriers occupy the
lowest Landau level, which results in a linear magnetoresistance in transport. Thus the linear MR observed by us is consistent with the presence of Dirac fermions with a linear energy dispersion.

From Fig.~\ref{fig:mr}(b) we can see that the transverse MR manifests a crossover at a critical field B$^{\star}$
from the semi-classical weak-field MR $\sim$ B$^{2}$ to the high-field
MR $\sim$ B dependence. The critical field
B$^{\star}$ above which quantum criterion is achieved is given as B$^{\star}$ = $\dfrac{1}{2e\hbar v_{F}^{2}}$ (E$_{F}$ + k$_{B}$T)$^{2}$.\cite{Huynh} Thus, if the magnetoresistance is dominated by Dirac Fermions with a linear dispersion, a parabolic dependence on $T$ is expected for the crossover field $B^*$.  Figure 5(c) shows the temperature dependence of B$^{\star}$ which can be fitted well with $\dfrac{1}{2e\hbar v_{F}^{2}}$ (E$_{F}$ + k$_{B}$T)$^{2}$. The fitting gives the values of v$_{F}$ $\sim$ 1.47 $\times$ 10$^{5}$ ms$^{-1}$ and E$_{F}$ $\sim$ 9.81 meV. The non-saturating linear magnetoresistance at high fields and low temperatures, and the $B^* \propto T^2$ presents strong evidence for the existence of Dirac Fermions in Pd$_{3}$Bi$_{2}$Se$_{2}$.

Kohler's rule states that for materials with one dominant scattering mechanism, the magneto-resistance (MR\%) in a magnetic field (B) can be represented in terms of scaling function F(x) as follows:\cite{Ziman}
\begin{equation}\label{key}$$
\centering
MR\% = F\Bigg($\dfrac{B}{\rho_{o}}$\Bigg) 
$$\end{equation}
where $\rho_{o}$ is the zero-field resistivity at a certain temperature. According to Kohler's rule, MR\% data is plotted as a function of B/$\rho_{o}$ as shown in Fig.~\ref{fig:mr}(d).
For Kohler's rule, MR\% vs. B/$\rho_{o}$ curves should collapse onto a single curve in a weak-field regime. Figure 5(d) shows the violation of Kohler's rule, which indicates the presence of more than one scattering mechanism.
Ahead of further discussions about magneto-transport in Pd$_{3}$Bi$_{2}$Se$_{2}$ thin films, we first investigate the dimensionality of thin films.
In the case of quantum interference (QI) effects, the relevant length scale is phase coherence length L$_{\phi} = \sqrt{D\tau}$, where $D$ is the diffusion constant and $\tau$ is the phase coherence time. Also, the criterion for the 2D nature of thin films is L$_\phi$ $>$ t, where 't' is the thickness of the film.\cite{Anderson,Ovadyahu} In our case, the value of L$_{\phi}$ is greater than the thickness of the film (based on the analysis below), indicating the 2D nature of thin films.

\begin{figure*}
	\centering
	\includegraphics[width=1\linewidth, height=0.5\textheight]{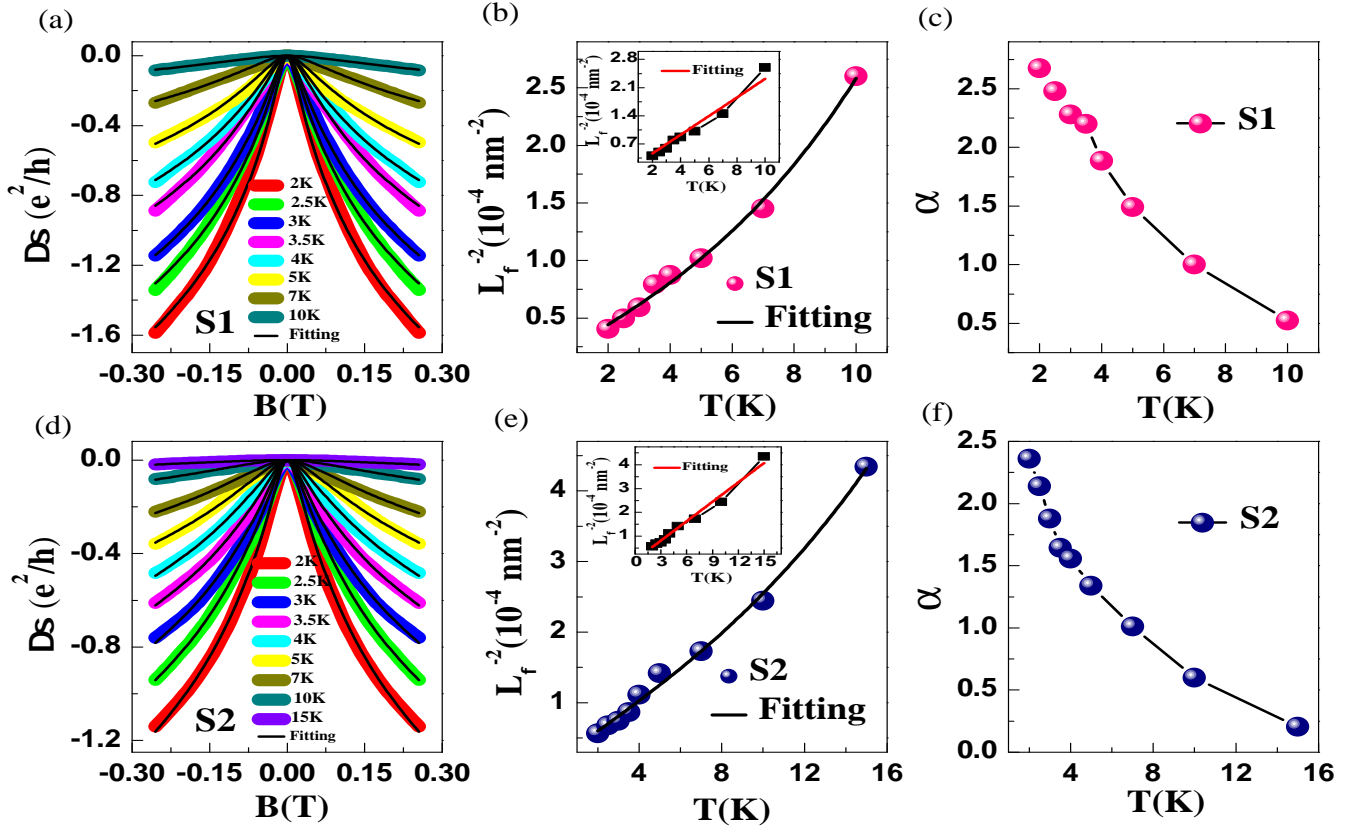}
	\caption{(a,d) shows the magnetic field dependence of magneto-conductance ($\triangle \sigma$) at various temperatures for S1 and S2. Black lines show fitted data w.r.t HLN equation. (b,e) Variation of L$_{\phi}$ as a function of temperature, revealing the contribution of different scattering mechanisms. (c,f) Temperature dependence of $\alpha$.}
	\label{fig:hln}
\end{figure*}

Figure~\ref{fig:hln}(a) shows the conductance correction ($\Delta \sigma$) vs. magnetic field at various temperatures in the range of $\pm$ 0.3 T for the S1 sample. Here, the two-dimensional conductance $\sigma_{xx}$ is obtained by inverting the MR tensor because the Hall resistance R$_{xy}$ of these films is much smaller than the longitudinal resistance R$_{xx}$. Therefore, we have $\sigma$(B) = (L/W) (1/R$_{s}$), L and W are the length and width of the sample, respectively, R$_{s}$ is the sheet resistance. For two-dimensional systems, Hikami-Larkin-Nagaoka (HLN) equation has been widely used to describe the effect of localization.\cite{Hikami,Fang} In the limit of high SOC expected for our material, Hikami-Larkin-Nagaoka (HLN) conductance correction can be written as
\begin{equation}\label{key}
\Delta \sigma (B)= -\alpha\dfrac{e^{2}}{\pi h}\bigg[\psi\bigg(\dfrac{1}{2} + \dfrac{B_{\phi}}{B} \bigg)- ln \bigg(\dfrac{B_{\phi}}{B}\bigg)\bigg], 
\end{equation}

where $\psi$ is the digamma function, B$_{\phi} = \hbar^{2}/(4eL_{\phi}^{2}$) is a characteristic field  with phase coherence length L$_{\phi} = \sqrt{D\tau}$.  $\alpha$ measures no. of conduction channels contributing to the transport. The parameter $\alpha$ is a prefactor expected to be 1/2 and -1 for weak anti-localization (WAL) and localization (WL), respectively.\cite{Hikami} However, $\alpha$ has been observed to deviate from 1/2 when more than one conducting channel contributes.\cite{Zhao,Shama} In particular, the value of $\alpha$ is 1/2 for one topological non-trivial conduction channel. If more than one conduction channel is present, each adds up and gives a value of $\alpha > 1/2$. \cite{Shama,Cao} 

Our data agree well with HLN fitting, as shown in Fig.~\ref{fig:hln}(a). The extracted values of $\alpha$ and L$_{\phi}$ as a function of temperature are shown in Fig.~\ref{fig:hln}(b-c). The value of $\alpha$ is 2.7 at $T = 2$~K, greater than the theoretical value expected for a single conduction channel, which implies the presence of more than one conducting topological channel. Figure~\ref{fig:hln}(c) shows the variation of $\alpha$ with temperature from 2.7 at 2K to 0.52 at 10 K, following the trend reported for several topological materials.\cite{Zhao,Shama,Cao}
Figure~\ref{fig:hln}(b) shows the temperature variation of L$_{\phi}$, which decreases from 157 nm at 2 K to 62 nm at 10 K.  The Nyquist electron-electron theory has predicted the temperature variation of L$_{\phi}$ as L$_{\phi} \propto T^{-n/2}$ where n = 1 for 2D systems.\cite{Wu} The inset of Fig. 6 (b) shows the T$^{-1/2}$ dependence of L$_{\phi}$, which indicates the failure of the fitting. This suggests that several scattering mechanisms are involved in dephasing the electron's phase. To analyze different scattering mechanisms, we use an equation which is given below :\cite{Bird}

 \begin{equation}\label{key}$$\centering
$\dfrac{1}{L_{\phi}^{2}}$ =$\dfrac{1}{L_{\phi o}^{2}}$ + A$_{ee}$ T$^{n}$ + B$_{ep}$ T$^{n'}$ 
$$\end{equation}
where $L_{\phi o}$ represents the zero temperature dephasing length, A$_{ee}$ T$^{n}$ and B$_{ep}$ T$^{n'}$ represent the contributions from electron-electron and electron-phonon interactions, respectively. According to electron-phonon interaction theory, electron-phonon interaction leads to n' $\ge$ 2.\cite{Sergeev,Reizer} We have obtained a proper fit by considering the electron-phonon interaction, as illustrated by the solid black line in Fig. 6 (b). The index n' varies between 2-3, which reveals the presence of e-p scattering mechanism, consistent with the observations in GeSb$_{2}$Te$_{4}$.\cite{Breznay}
The value of A$_{ee}$ = 1.57 $\times$ 10$^{-5}$ is large in comparison to the  B$_{ep}$ = 8.90 $\times$ 10$^{-8}$, which suggests the dominance of the e-e scattering mechanism.

Figure 6 (d) shows the magnetic field dependence of conductance at various temperatures in the range of $\pm$ 0.25 T for the S2 sample. By applying the eq. 4, we have done fitting of magneto-conductance and extracted the parameters L$_{\phi}$ and $\alpha$, as shown in fig. 6 (e-f). Figure 6 (e) shows the temperature dependence of L$_{\phi}$, which decreases from 133 nm at 2 K to 48 nm at 15 K. Similar to the S1 sample, we have done the linear fitting of L$_{\phi}$$^{-2}$, which indicates the failure of the fitting. It suggests that several scattering mechanisms are involved in dephasing electron's phase. By considering the electron-phonon interaction, we have obtained a proper fit with n' = 3, as illustrated by the solid black line in Figure 6(e), which reveals the presence of an e-p scattering mechanism.\cite{Sergeev,Reizer} The value of A$_{ee}$ = 2.05 $\times$ 10$^{-5}$ is large as compared to the value of B$_{ep}$ = 3.18 $\times$ 10$^{-8}$, which suggests the dominance of the e-e scattering mechanism.  The value of $\alpha$ is 2.36 at T = 2 K, which is greater than the expected value. The  $\alpha$ $>$ 0.5 indicates the presence of more than one topological non-trivial conduction channels.\cite{Cao}
 
\section{Conclusion}
We report successful synthesis of thin films of the topological material Pd$_{3}$Bi$_{2}$Se$_{2}$.  We have demonstrated that the linear energy dispersion causes unusual non-saturated linear magnetoresistance coming from the Dirac fermions inhabiting the lowest Landau level in the quantum limit. At a critical field B$^{\star}$, the transverse magnetoresistance shifts from semi-classical weak-field B$^{2}$ to high-field B dependence. The critical field B$^{\star}$ follows a quadratic temperature dependence, which is also expected from the Landau level splitting of a linear energy dispersion. Thus our results evince the existence of Dirac Fermions in Pd$_{3}$Bi$_{2}$Se$_{2}$.  We also observed signatures of weak anti-localization (WAL). The magnetoconductivity data are satisfactorily analyzed in terms of the Hikami-Larkin-Nagaoka theory. It was found that the coefficient $\alpha$ (number of conduction channels) deviates from the value 0.5 expected for 2D systems with a single topological conduction channel. This indicates the contribution from additional topological conducting channels in the electron transport. Dependence of the dephasing length L$_{\phi}$ on temperature is also anomalous. We found that this behavior can be understood by including both the Nyquist electron-electron scattering and electron-phonon scattering as the phase relaxation mechanism in PBS films. The magnetoresistance data show deviations from Koehler’s rule, suggesting multiple scattering mechanisms. Finally, the resistance versus temperature data show signatures consistent with a CDW transition. Such signatures have not been reported in previous works on bulk samples. These anomalous behaviors make Pd$_{3}$Bi$_{2}$Se$_{2}$ an interesting system for further study in various morphologies.

\section{acknowledgment}
We acknowledge use of the central x-ray and SEM facilities at IISER Mohali.  We acknowledge use of the PPMS Sonipat (CFMS 14T, Cryogenic Ltd.), Central Research Facility, IIT Delhi for some transport measurements.

\end{document}